\documentclass[twocol]{ametsocV5}

\usepackage{amsmath,amsfonts,amssymb,bm}
\usepackage{mathptmx}
\usepackage{newtxtext}
\usepackage{newtxmath}
\usepackage{multirow}

\usepackage{lipsum}

\title{Seamless multi-model postprocessing for air temperature forecasts in complex topography}

\authors{Regula Keller\correspondingauthor{Regula Keller, regula.keller@env.ethz.ch \newline \newline \textbf{Notice} \newline This work has not yet been peer-reviewed and is provided by the contributing authors as a means
to ensure timely dissemination of scholarly and technical work on a noncommercial basis. Copyright
and all rights therein are maintained by the authors. It is understood that all persons copying this
information will adhere to the terms and constraints invoked by each author's copyright. This work
may not be reposted without explicit permission of the copyright owner.}} 

\affiliation{Federal Office of Meteorology and Climatology MeteoSwiss, Zurich, Switzerland, and Centre for Climate Systems Modelling (C2SM), ETH Zurich, Zurich, Switzerland, and Institute for Atmospheric and Climate Science, ETH Zurich, Zurich, Switzerland}

\extraauthor{Jan Rajczak, Jonas Bhend, Christoph Spirig, Stephan Hemri, Mark A. Liniger}
\extraaffil{Federal Office of Meteorology and Climatology MeteoSwiss, Zurich, Switzerland}

\extraauthor{Heini Wernli}
\extraaffil{Institute for Atmospheric and Climate Science, ETH Zurich, Zurich, Switzerland}

\abstract{Statistical postprocessing is routinely applied to correct systematic errors of numerical weather prediction models (NWP) and to automatically produce calibrated local forecasts for end-users. Postprocessing is particularly relevant in complex terrain, where even state-of-the-art high-resolution NWP systems cannot resolve many of the small-scale processes shaping local weather conditions. In addition, statistical postprocessing can also be used to combine forecasts from multiple NWP systems. Here we assess an ensemble model output statistics (EMOS) approach to produce seamless temperature forecasts based on a combination of short-term ensemble forecasts from a convection-permitting limited-area ensemble and a medium-range global ensemble forecasting model. We quantify the benefit of this approach compared to only processing the high-resolution NWP. We calibrate and combine 2-m air temperature predictions for a large set of Swiss weather stations at the hourly time-scale. The multi-model EMOS approach ('Mixed EMOS') is able to improve forecasts by 30\% with respect to direct model output from the high-resolution NWP. A detailed evaluation of Mixed EMOS reveals that it outperforms either single-model EMOS version by 8-12\%. Valley location profit particularly from the model combination. All forecast variants perform worst in winter (DJF), however calibration and model combination improves forecast quality substantially.}

\begin{document}

\maketitle
%
%
%

%

\section{Introduction}
Weather forecasts are a key element to support decision-making for a broad range of applications. Thus there is a high demand for accurate weather forecasts from a wide range of stakeholders including the general public, the private sector, and authorities issuing weather warnings. Over the last decades, forecasts have been steadily improving largely driven by advances in numerical weather prediction (NWP) including the data assimilation procedure \citep{Bauer_2015}. 
The advent of ever more powerful high performance computers allows simulating weather with increasing detail. In addition, multi-model ensemble prediction systems are run to quantify the uncertainty of forecasts. Despite these improvements in NWP, forecasts from physics-based models are not free from systematic bias, and ensemble predictions are often underdispersive \citep{Wilks_2018_ch1}. At the same time, the rapidly increasing data volume produced by state-of-the-art NWP systems poses a significant challenge to end users aiming for accurate and easily interpretable products. Furthermore, users usually require one forecast but they may have to choose between different models depending on the time horizon of the forecast they need. And a last challenge is the availability of short-range predictions and the time it takes the users to receive and evaluate them for their specific purpose; therefore often, users cannot profit from short-range high-resolution forecasts. 
\\

Statistical postprocessing is an attractive tool to further refine, improve and calibrate NWPs and at the same time generate end-user tailored products. The principle of statistical postprocessing is to describe empirical relationships and (or) error-characteristics of past forecast-observation pairs, which are then used to correct the most recent forecasts. The goal of statistical postprocessing is to maximize sharpness subject to calibration \citep{Gneiting_2007}. The pioneering work on so-called model output statistics (MOS) goes back to \citet{Glahn_1972} and was applied successfully in e.g., the Netherlands for improving various forecast parameters including maximum and minimum temperature \citep{Lemke_1988}. Since then, MOS and other postprocessing methods have become increasingly popular and many different approaches and variants have been proposed for deterministic and probabilistic forecasts \citep{Vannitsem_2018}. Non-parametric ensemble postprocessing often considers quantiles \citep{Bremnes_2004,Taillardat_2016}. The most common parametric ensemble postprocessing methods are Bayesian model averaging \citep[BMA;][]{Raftery_2005} and ensemble model output statistics \citep[EMOS;][]{Gneiting_2005}. A further EMOS-related method is Standardized Anomaly Model Output Statistics \citep[SAMOS;][]{Dabernig_2016}. Recent research efforts increasingly exploit machine-learning approaches, such as neural networks \citep{RaspLerch_2018}.  
EMOS has been found to be a simple yet skillful approach to postprocess ensemble forecasts enabling the generation of calibrated probabilistic forecasts \citep{Gneiting_2005}. The principle of EMOS is a regression that corrects for errors in the mean (e.g., systematic biases) and spread (e.g., under- or overdispersion) of an ensemble forecast. In analogy to multiple regression, EMOS offers flexibility to be extended to a multi-predictor and (or) multi-model framework. 
\\

This study explores a high- and a coarse-resolution NWP ensemble and their combination, and investigates the accuracy of probabilistic 2-m air temperature forecasts at measurements stations spread across Switzerland. High-resolution NWP is essential to provide forecasts of local weather up to a few days ahead, especially in regions with complex topography as the Alps. Global, coarser-resolution NWP provide longer-range forecasts. Here, we implement a straight-forward EMOS approach to calibrate the two NWP ensembles and to combine the information from both NWP models into a single data stream. Typically, multi-model ensemble calibration, weighting, and combination are done in separate steps \citep[e.g.,][]{Johnson_2009,Beck_2016}. Our approach performs these different tasks in one step and includes a seamless transition between multi-model and single-model prediction beyond the forecast horizon of the high-resolution ensemble.
Often, simply adding a longer-rage forecast at the end of a shorter-range forecast is already called 'seamless' \citep{Wastl_2018,Wetterhall_2018}. Here, we propose two simple methods to smooth this transition from multi-model to single-model prediction. 
\\

The paper is structured as follows: Section \ref{sec:data} provides an overview of the data and Section \ref{sec:meth} describes the methods used in this study, then the results are presented in Section \ref{sec:resu}, and finally Section \ref{sec:disc} presents a discussion and the conclusions.  

\section{Data}\label{sec:data}

\subsection{Observational Reference}
The observational reference in this study are 2-m air temperature measurements from 290 sites in Switzerland (see Fig. \ref{fig:map}a). The majority of these automatic measurement stations are operated by the Swiss Federal Office of Meteorology and Climatology (MeteoSwiss). These observational data are from high-quality instruments and have gone through extensive quality control (automatic and manually). The data-set also includes measurements form several partner networks operated by public authorities, research institutes, and private weather services. The quality of these measurements is lower, as partner data stems from various instrument (mostly high quality) and has been subject to only basic quality control. Nevertheless, our forecasts show comparable scores for data from both origins, hence both are used in this study. The complete station set includes a large variety of locations within the pre-Alpine lowlands as well as in topographically highly complex settings within the Alps, including valley-floor and mountain-top stations. The majority of observations come form below 1000 m (55\%), with the lowest at 200 m, and the highest at 3571 m.

\subsection{Numerical Weather Prediction Models}
\begin{figure*}[t]
  \noindent\includegraphics[width=39pc,angle=0]{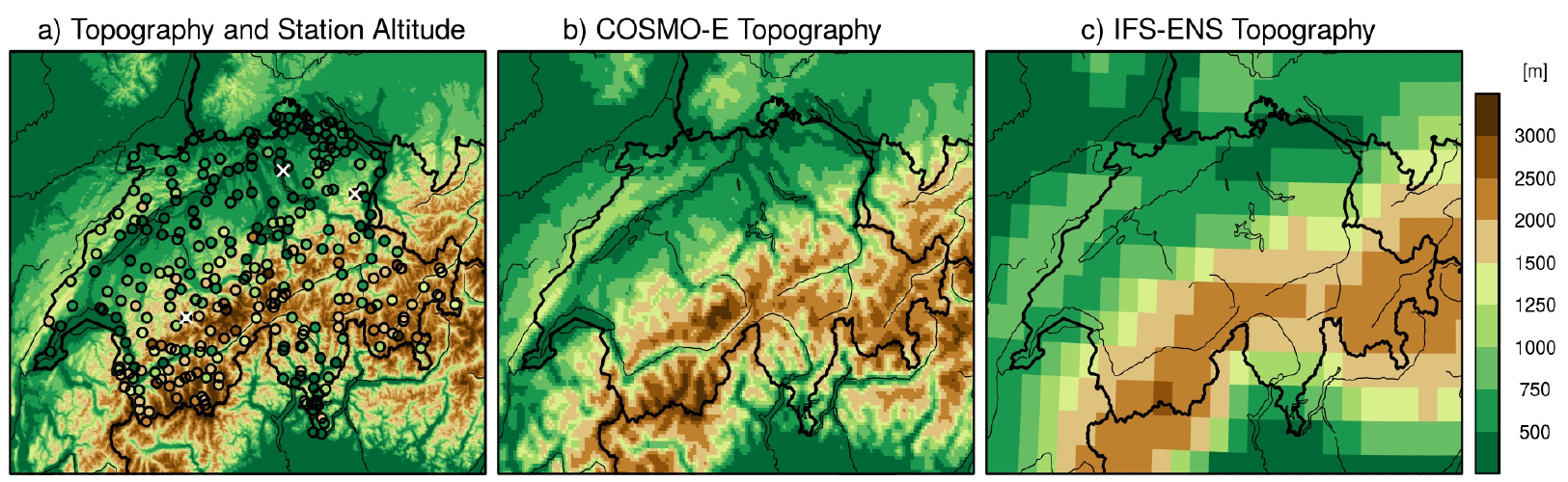}\\
  \caption{a) Topography of Switzerland at 500 m resolution. The circles mark the location of the stations and their elevation is indicated by color. The white crosses mark (from left to right) Adelboden, Z\"urich, and S\"antis. b) Model topography of COSMO-E in and around Switzerland. c) Same as b) for IFS-ENS.}
  \label{fig:map}
\end{figure*}

Two state-of-the-art operational NWP ensembles are used in this study, which are based on the high-resolution numerical weather prediction model from the Consortium for Small-scale Modeling (COSMO-E) operated by MeteoSwiss, and on the coarser-resolution Integrated Forecasting System (IFS-ENS) from the European Centre for Medium-Range Weather Forecasting (ECMWF). IFS-ENS is a 51-member global ensemble at about 18 km horizontal resolution \citep{IFS}. It is initialized four times daily with forecasts out to 6 days (initializations at 06 and 18 UTC) and 15 days (initializations at 00 and 12 UTC). COSMO-E is a limited-area model with 2.2-km grid spacing for the greater Alpine region offering twice a day (00 and 12 UTC) a set of 21 members with forecasts extending to 5 days \citep{COSMO}. At the lateral boundaries, COSMO-E forecasts are forced by the 18 and 06 UTC IFS-ENS simulations, respectively. Figures \ref{fig:map}b and \ref{fig:map}c visualize the representation of the topography of Switzerland in both models. This study relies on an archive of the operational 00 and 12 UTC runs from both ensemble systems for the time period from 01 January 2017 until 27 October 2019 (i.e., 2 years and 300 days). 
\\

Analyses have been carried out for 00 and 12 UTC runs separately and show consistent results; thus this paper will merely focus on the results for the 00 UTC model runs. The focus is on a 120-h forecast horizon, which is the time period covered by both models, and the few following hours for the transition from a multi-model to a single-model system. As the availability of the IFS-ENS output changes from 1-hourly to 3-hourly timesteps after a lead time of 90 h, hourly timesteps have been obtained by linear interpolation. A detailed overview of the model attributes is given in Table \ref{tab:models}. In the following, COSMO-E is called COSMO and IFS-ENS just IFS.

\section{Methods}\label{sec:meth}

\subsection{Ensemble Model Output Statistics}
Statistical postprocessing aims at correcting systematic biases in NWP output. We here consider the well-established EMOS methodology, also termed non-homogeneous Gaussian regression \citep{Gneiting_2005}. EMOS allows to calibrate probabilistic forecasts by correcting for errors in the mean and variance. An EMOS forecast is characterized by the parameters of a probability density function (PDF). The PDF should best match distributional characteristics of the predictand. In the case of 2-m air temperature, a Gaussian distribution is used \citep{Gneiting_2005,Scheuerer_2014}. Other variables may require different distributional assumptions \citep{Wilks_2018}. 
\\

In a very straightforward setup, we use as parameters for our Gaussian predictive PDF the ensemble mean 
\begin{equation}
    \mu(t) = a + b \overline{x(t)}
    \label{eq:emos_mean}
\end{equation}
and ensemble standard deviation 
\begin{equation}
    \sigma(t) = \sqrt{c^2 + d^2 s^2(t)},
	\label{eq:emos_std}
\end{equation}
where $\overline{x(t)}$ and $s(t)$ denote the mean and the standard deviation of the direct model output (DMO) at time $t$, respectively, and $a$, $b$, $c$ and $d$ are the regression coefficients. Following the approach of \citet{Gneiting_2005} the regression coefficients are estimated by minimizing the continuous ranked probability score (CRPS, see below). 
\\

The coefficients are estimated, separately for each location and lead time, by using a rolling archive (i.e., training period) that incorporates the past 45 days. The choice of using 45 days has operational advantages because models change relatively frequently ans such a short training period guarantees that training is done only with the current or very recent versions. Also it allows to partly consider seasonality including seasonally-specific weather types, but it can be prone to errors in the case of abrupt changes in weather conditions, in particular during transition seasons and in the case of snow-cover retreat in the Alpine region. Using reforecasts could reduce these errors \citep{Wilks_2007,Hagedorn_2008}, however this usually not operationally feasible. Sensitivity tests (not shown) with different lengths of rolling archive indicated that a window of 45 days is a good choice for 2-m air temperature, in agreement with previous findings \citep{Gneiting_2005,Hagedorn_2008}.

\subsection{Multi-Model Combination}
In order to merge two NWP systems, the EMOS equations are extended such that they allow incorporating an additional predictor for the mean and the standard deviation:
\begin{equation}
\begin{split}
    \mu(t) = a + b_1 \overline{x_1(t)} + b_2 \overline{x_2(t)} \\
    \sigma(t) = \sqrt{c^2 + d^2_1 s^2_1(t) + d^2_2 s^2_2(t)},
\end{split}
\end{equation}
where $\overline{x_1(t)}$ and $s^2_1(t)$ are the mean and standard deviation of the raw COSMO forecast, and $\overline{x_2(t)}$ and $s^2_2(t)$ the mean and standard deviation of the raw IFS forecast. This combination is termed Mixed EMOS in the following. In this study, we combine two models only, but the approach could easily be extended to include more models. 
\\

Coefficients $b_1$, $b_2$, $d_1$ and $d_2$ are constrained to be positive so the weight of a single predictor (for instance, the "importance" of a NWP model ensemble) can be determined as follows: 
\begin{equation}
\begin{split}
    Weight\ for\ Mean = \frac{b_1}{(b_1 + b_2)} \\
    Weight\ for\ Standard\ Deviation = \frac{d_1}{(d_1 + d_2)},
\end{split}
\end{equation}
where $Weight\ for\ Mean$ and $Weight\ for\ Standard\ Deviation$ is the weight of predictor 1. In the present case, COSMO serves as predictor 1 and is compared to its fractional weight with respect to predictor 2, which is the IFS. 
\\

\subsection{Transition}
The forecast period of Mixed EMOS is limited by the maximum common lead time of both models. In our case, the forecast period of COSMO is 120 h, while IFS extends up to 360 h. Thus Mixed EMOS can only be applied up to lead time 120 h, and thereafter, forecasts can only be based on IFS. In order to facilitate a smooth transition between Mixed EMOS and IFS EMOS we test two approaches. While the first approach modifies the predictions during the three hours before the transition, the second approach affects the three following timesteps.
\\

For the first approach (referred to as transition 1), upper bounds are defined for coefficients $b_1$ and $d_1$ to limit the weight of COSMO. For lead time $t = 118,\ 119,\ 120$ h, the upper bounds are defined as 
\begin{equation}
    b_1(t)=b_1(117\ \textrm{h}) w(t),
\end{equation}
and analogously for $d_1$, where the weights decrease linearly with lead time and attains the values 0.75, 0.5, 0.25 for $t = 118,\ 119,\ 120$ h, respectively. The second approach (transition 2) prolongs the influence of COSMO to the three hours beyond its forecast horizon. The difference of $\mu(120\ \textrm{h})$ between Mixed EMOS and IFS EMOS is taken and then added to the following three hours with decreasing weights, defined as above. The same is done for $\sigma(120\ \textrm{h})$.
\\

We only present results from the transition of the 00 UTC runs for consistency but also because, in our setup, it is the more complex transition. The 00 UTC run transitions from multi-model to single-model at midnight, i.e., at a time of the day when the model that runs out (COSMO) is typically more important than the continuing IFS - as discussed below.

\subsection{Validation}
The grid point of the NWP model that is nearest to the station is used for comparison with the station observations. However, this model grid point might be at a different altitude than the station. It is to be noted that even the 2.2-km NWP system COSMO is not able to fully resolve the complex topography in Switzerland (see also Fig. \ref{fig:map}). For instance, the largest mismatch of model topography against target elevation of the corresponding station in the present study is 934 m for the 2.2-km COSMO, and 1687 m for the 18-km IFS. 
To enable a fair comparison between the DMO and EMOS-forecasts, DMO is corrected for its altitudinal offset with respect to target locations using a constant lapse rate correction of 0.6\textdegree C per 100 m.
\\

The performance of different forecasts is assessed using the continuous ranked probability score \citep[CRPS;][]{Gneiting_Raftery_2007}. The CRPS is a proper scoring rule and a common measure to evaluate probabilistic forecasts. It takes the integrated squared difference between the cumulative distribution functions (CDFs) of the forecast and the observation. Therefore, smaller values indicate more accurate predictions and a value of 0 denotes a perfect forecast.
\\

In order to directly compare different strategies, we use the continuous ranked probability skill score (CRPSS):
\begin{equation}
	CRPSS = 1 - \frac{CRPS}{CRPS_{ref}},
\end{equation}
where $CRPS$ is the mean score of the forecast under investigation and $CRPS_{ref}$ denotes the score of a reference forecast. A positive CRPSS stands for an improvement of the forecast compared to the reference, with a maximum value of 1 for a perfect forecast. Negative values indicate weaker performance than the reference. 
To assess the significance of differences in the score between different models (Fig. \ref{fig:day-night}), we use the Diebold-Mariano test \citep{Diebold_1995} as implemented in the R package 'SpecsVerification' \citep{SpecsVerification}.
\\

In addition, we evaluate the reliability of the forecast. Reliability describes the degree to which the forecast probability and the observed frequencies agree \citep[e.g.,][]{Weigel_2012}. We analyze the reliability of the different forecasting strategies using probability integral transform (PIT) histograms, which show the relative position of the observation within the ensemble distribution, summed up over many forecasts. Ideally, the histogram has a uniform shape. One-sided PIT histograms indicate a bias, and an u- or n-shape reveals underdispersion or overdispersion, respectively.
\\

The topographic position index (TPI) is used to characterize the topographic situation of the investigated sites and to asses the impact of topography on the results (Figs. \ref{fig:day-night} and \ref{fig:DJF}). The TPI corresponds to the difference in elevation of a central pixel relative to the mean altitude of its surrounding eight pixels. A positive TPI indicates an elevated position, e.g., a mountain top or ridge, and a negative TPI denotes a cavity such as a valley. The TPI depends on the grid spacing of the topography dataset; we calculate the TPI with a 500-m resolution topography, as this has been found to characterize local-scale conditions fairly well (not shown).

\section{Results}\label{sec:resu}

\subsection{Verification of Forecasts}
The mean forecast score CRPS as a function of lead time is presented in Fig. \ref{fig:CRPS}a for all analyzed forecast variants. Scores for the elevation-corrected raw model output from COSMO and IFS, their postprocessed EMOS counterparts, as well as the combined Mixed EMOS are shown in a comparative manner. 
The forecast quality of both NWP models varies strongly with time of day. While IFS performs best in the morning (recall that lead time 0 corresponds to 00 UTC, which is local time 1 h in winter and 2 h in summer), COSMO tends to have the best score in the evening. COSMO is worst in the early afternoon, the only time of the day when it is on average outperformed by IFS. The application of EMOS lowers the CRPS of COSMO and IFS at all lead times. Both postprocessed forecasts still exhibit the same diurnal cycles but less prominently so. With EMOS, the best performance of IFS shifts to the evening. The discrepancy in forecast quality between the two models has clearly decreased with EMOS postprocessing. The Mixed EMOS outperforms both single-model EMOS at all lead times. The diurnal cycle of the Mixed EMOS score is highly correlated with the score of COSMO EMOS but at lower CRPS values. As expected, the CRPS tends to increase with lead time, but this increase is weak compared to the diurnal variations.
\\

The seasonal score CRPS and skill score CRPSS of the different forecast approaches are summarized in Table \ref{tab:scores}. The skill score is calculated with the elevation-corrected COSMO as reference. The high-resolution NWP model COSMO outperforms the coarser-resolution IFS in all seasons, particularly during winter (DJF) when forecast quality is lowest for both models. EMOS is able to improve the forecast of COSMO in a distinct manner by 24\% and of IFS by 20\%. In the case of COSMO, EMOS improves the skill at all stations in spring and winter, and at more than 98\% of the stations in summer and fall. Forecast quality of IFS EMOS improves for more than 90\% of cases at the annual scale. Seasonally stratified, values range from 73\% in fall to 89\% in spring. Also with EMOS, both models still exhibit weakest performance during winter. COSMO EMOS and IFS EMOS have similar quality in spring and summer. In fall and winter, COSMO EMOS clearly outperforms IFS EMOS. 
\\

The Mixed EMOS improves forecasts by $\sim$30\% with respect to elevation-corrected COSMO, by 8.8\% with respect to COSMO EMOS and by 12\% with respect to IFS EMOS. The Mixed EMOS outperforms all other forecasts in all seasons. The general performance is best in summer (CRPS of 0.978), followed by fall (1.02), spring (1.06) and winter (1.17). All five forecasting variants under investigation share the same seasonal rank-order in terms of forecast quality. The improvement in terms of CRPSS of Mixed EMOS over the operationally used weather forecasting model COSMO is largest in spring and winter. The benefit of Mixed EMOS is present in all stations in all seasons except summer, when one single station has a CRPSS of $\sim$0. 
\\

Location-specific analyses for three geographically diverse stations are presented in Fig. \ref{fig:CRPS}b-d in terms of annual mean CRPS as a function of lead time. The three stations are used to illustrate three characteristic examples of locations in Switzerland and enable to study model behavior in detail. Z\"urich is a typical low-land pre-Alpine location at 556 m MSL, S\"antis is a mountain-top station at 2501 m MSL that distinctly stands out into the free atmosphere, and Adelboden an Alpine valley floor location at 1321 m MSL prone to local-scale processes like cold-pool formation, shading effects and local-scale wind-systems. It should be noted that valley stations are an inhomogeneous group and Adelboden is not representative for all of them.
\\

It is obvious in Fig. \ref{fig:CRPS}b that for Z\"urich both models already have good scores; only IFS stands out negatively at night. EMOS can improve this shortcoming and slightly improve both forecasts at all lead times. The different EMOS variants exhibit very similar scores. At all times of day, the Mixed EMOS narrowly outperforms either single-model EMOS for Z\"urich.
For S\"antis (Fig. \ref{fig:CRPS}c), IFS has distinct biases (CRPS of 3.17). This is primarily because of the horizontal and vertical mismatch between the target location and the coarse resolution of IFS topography and is to some extent also evident for COSMO. IFS is also characterized by a a large diurnal cycle in score with weakest performance in the early morning and best performance during the afternoon. EMOS is able to massively improve air temperature predictions at S\"antis, particularly for IFS. The Mixed EMOS is mostly consistent with COSMO EMOS, but performs slightly better on average.
In Adelboden (Fig. \ref{fig:CRPS}d) both models show mediocre scores as compared to the entire set analyzed. Postprocessing with EMOS (and model combination) is able to remove large parts of these deficits but is not able to reach the low CRPS levels seen at low-land and mountain-top locations. Still, Adelboden is the location where the benefit of the Mixed EMOS is largest due to important shortcomings in both models.
\\

An in-depth analysis of the benefit of Mixed EMOS with respect to COSMO and its post-processed counterpart during day (07-18 UTC) and night (19-06 UTC) is shown in Fig. \ref{fig:day-night}. 
During the day, the gain in forecast quality of Mixed EMOS relative to COSMO is most prominent in elevated valleys (altitude of 1000-2000 m and negative values of TPI). 
At night, the skill of Mixed EMOS cannot be related to the geographic characteristics as easily. Largest improvements of Mixed EMOS occur in the Alps albeit with a lot of variability between stations. Skill at stations in the Swiss plateau (characterized by altitude around 500 m and north of 47\textdegree N with small TPI magnitude) is generally limited to 10-20\%.
It is, however, interesting to note that Mixed EMOS performs significantly better than COSMO EMOS for all sites but three during the day. While multi-model combination improves forecast quality at most stations during the night, at exposed locations (i.e., elevated stations with positive TPI) along the northern slopes of the Alps, the Mixed EMOS forecasts perform significantly worse than COSMO EMOS.
\\

\begin{figure}[t]
  \noindent\includegraphics[width=19pc,angle=0]{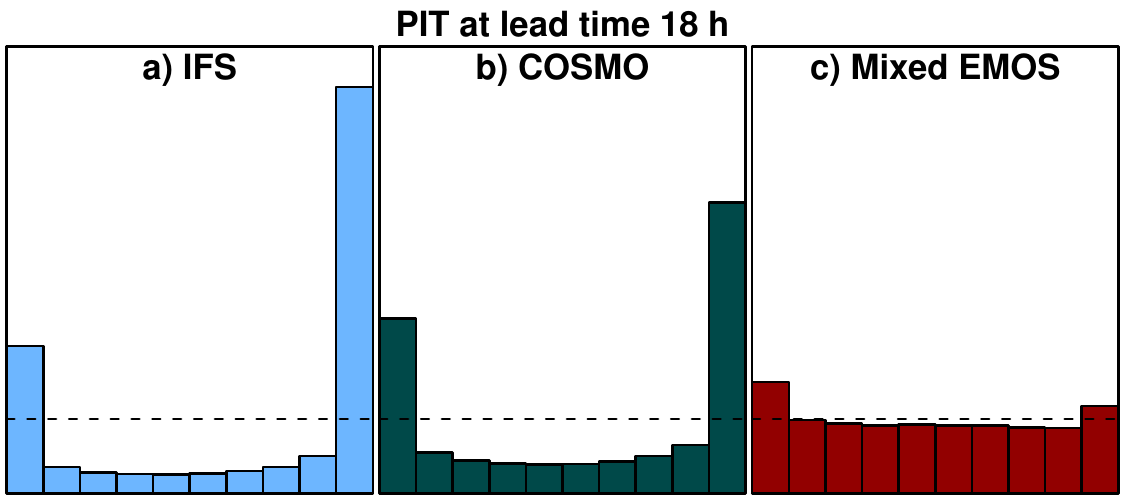}\\
  \caption{PIT diagram of elevation-corrected IFS and COSMO, and Mixed EMOS at lead time 18 h.}
  \label{fig:pit}
\end{figure}

The reliability of the different forecasting strategies is analyzed using PIT histograms. Figure \ref{fig:pit} shows a PIT histogram for lead time 18 h (i.e., in the early evening; other lead times are not shown as they show similar results). Both, COSMO and IFS are strongly underdispersed and possess a systematic cold bias, in particular the IFS. PIT diagrams at other lead times show similar results; only around noon does COSMO show no or a small positive bias and then the negative bias in IFS is the smallest.
Applying the Mixed EMOS reduces underdispersion distinctly (the same is valid for single-model EMOS using COSMO and IFS; not shown), which indicates an improved reliability of EMOS-postprocessed temperature predictions.

\subsection{Model Weights in Mixed EMOS}
\begin{figure*}[t]
  \centerline{\noindent\includegraphics[width=33pc,angle=0]{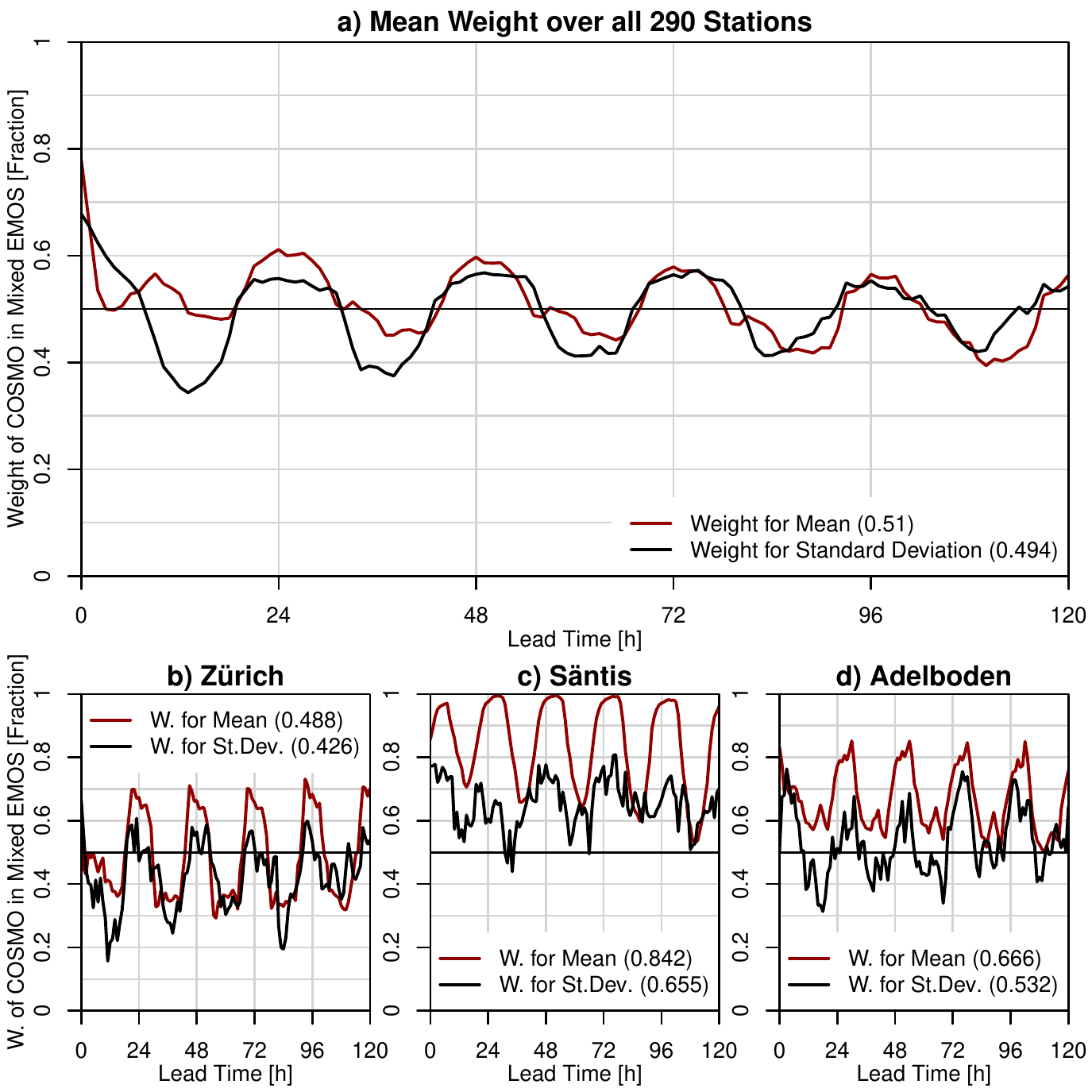}}
  \caption{The weight of COSMO direct model output in the Mixed EMOS for the mean (red lines) and standard deviation (black lines). Averages over all dates and a) all stations, b) Z\"urich (lowland), c) S\"antis (mountain top), and d) Adelboden (valley floor). The values in brackets are averages over the first 120 lead times. A value > 0.5 means that COSMO has higher weight than IFS in the Mixed EMOS.}
  \label{fig:wght}
\end{figure*}

Figure \ref{fig:wght}a depicts the 290-station average weight of COSMO, $w$, as a function of lead time. The weight of IFS consequently is $1-w$. The weights for the mean and for the standard deviation show a similar pattern. While IFS is more important (i.e., $w < 0.5$) during day-time and towards longer lead times, COSMO is the favoured model during the night and at shorter lead times, especially during the first few hours. The times when IFS has more weight correspond to the cases when IFS EMOS outperforms COSMO EMOS (see Fig. \ref{fig:CRPS}a). The weight of COSMO for the standard deviation is generally lower (i.e., IFS is more important), especially on day one. Both raw models are underdispersed, but IFS less so, therefore has a higher weight for the spread.
\\

Figures \ref{fig:wght}b-d illustrate the model weighting (i.e., importance of COSMO at given locations) for the three exemplary locations discussed earlier (see Figs. \ref{fig:CRPS}b-d). For the two stations located within complex topography (S\"antis and Adelboden) COSMO is the model in favour compared to IFS.
At the station of Z\"urich in the lowland the average weight of COSMO is 0.49 for the mean and 0.43 for the standard deviation, indicating an equal importance of the two models overall. However, the weight has a distinctive diurnal cycle with high values ($w >0.6$) during the night when COSMO has a better score, and low values at day-time ($w < 0.4$) when IFS is the preferred model.
The mountain station S\"antis has high weights for COSMO (0.84 and 0.66 for the mean and the standard deviation, respectively), with a strong diurnal cycle (weight for mean near 1 at night and 0.5-0.7 during the afternoon). In the case of the valley-floor location Adelboden, the weight of COSMO is at roughly 0.67 for the mean and 0.53 for the standard deviation.
\\

\subsection{Transition}
\begin{figure*}[t]
  \centerline{\noindent\includegraphics[width=33pc,angle=0]{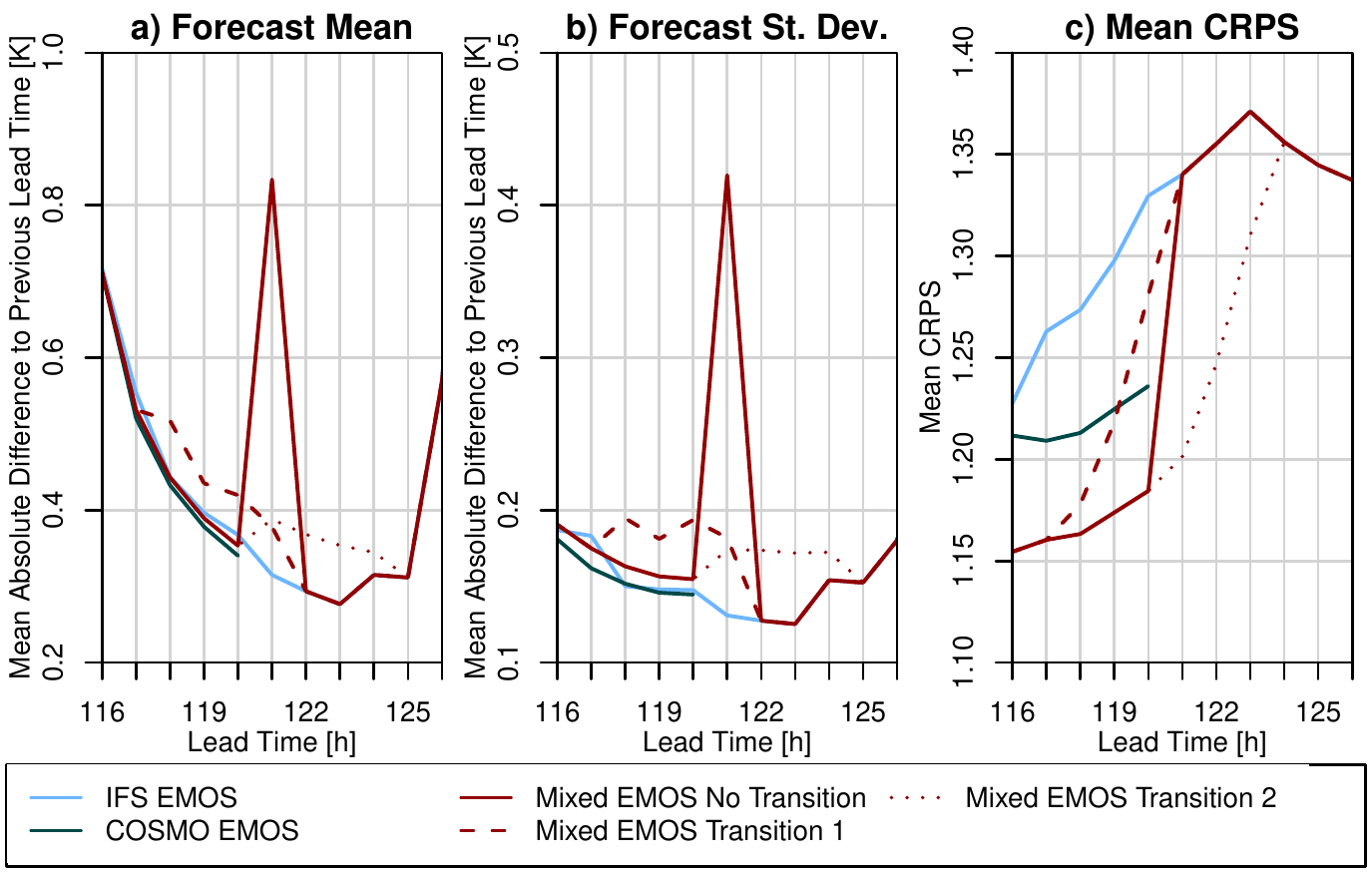}}
  \caption{a) The mean absolute difference to the previous lead time of the forecast mean, b) the mean absolute difference to the previous lead time of the forecast standard deviation, and c) the mean CRPS during the transition period from forecast hours 116 to 126 for IFS EMOS, COSMO EMOS, Mixed EMOS without any transition, and Mixed EMOS with transitions 1 and 2 (see text for explanation).}
  \label{fig:trans}
\end{figure*}

Figure \ref{fig:trans} shows the forecast score and 'smoothness' in the transition phase from a multi-model to a single-model forecast. The 'smoothness' is displayed as the mean absolute difference in the forecast between the forecast at the time of interest and one hour earlier (i.e., the previous lead time). Both for predicted mean temperature (Fig. \ref{fig:trans}a) and its standard deviation (Fig. \ref{fig:trans}b) the transition between the forecasting strategies is substantial, with a spurious peak in short-term changes, in the case where no specific transition smoothing is applied. The two simple transition approaches applied here (i.e., transition 1 and transition 2) provide 'smooth' forecasts but still possess a signal with respect to the continuous (single-model) IFS EMOS. Note that due to the diurnal cycle of the temperature forecast, there is always a difference to the forecast at the previous lead time.
\\

The CRPS score of the different transition methods is presented in Fig. \ref{fig:trans}c. Obviously transition 2 is the preferable approach as the overall performance is best. Transition 1 decreases the forecast quality in the last hours before COSMO fades out, while transition 2 increases the quality beyond the transition at 120 h and even improves predictions in the following hours. 

\section{Discussion and Conclusions}\label{sec:disc}
The present study demonstrates that statistical post-processing and model combination is able to distinctly improve 2-m air temperature forecasts of state-of-the-art operational NWP systems. Using the EMOS method, we show that postprocessing removes substantial parts of the systematic biases inherent to weather prediction models and further adjusts for errors in ensemble spread and thereby improves reliability. We particularly assessed differences in performance of the global coarser-resolution IFS and the regional higher-resolution COSMO ensembles and their combination.
\\

\begin{figure*}[t]
  \centerline{\noindent\includegraphics[width=33pc,angle=0]{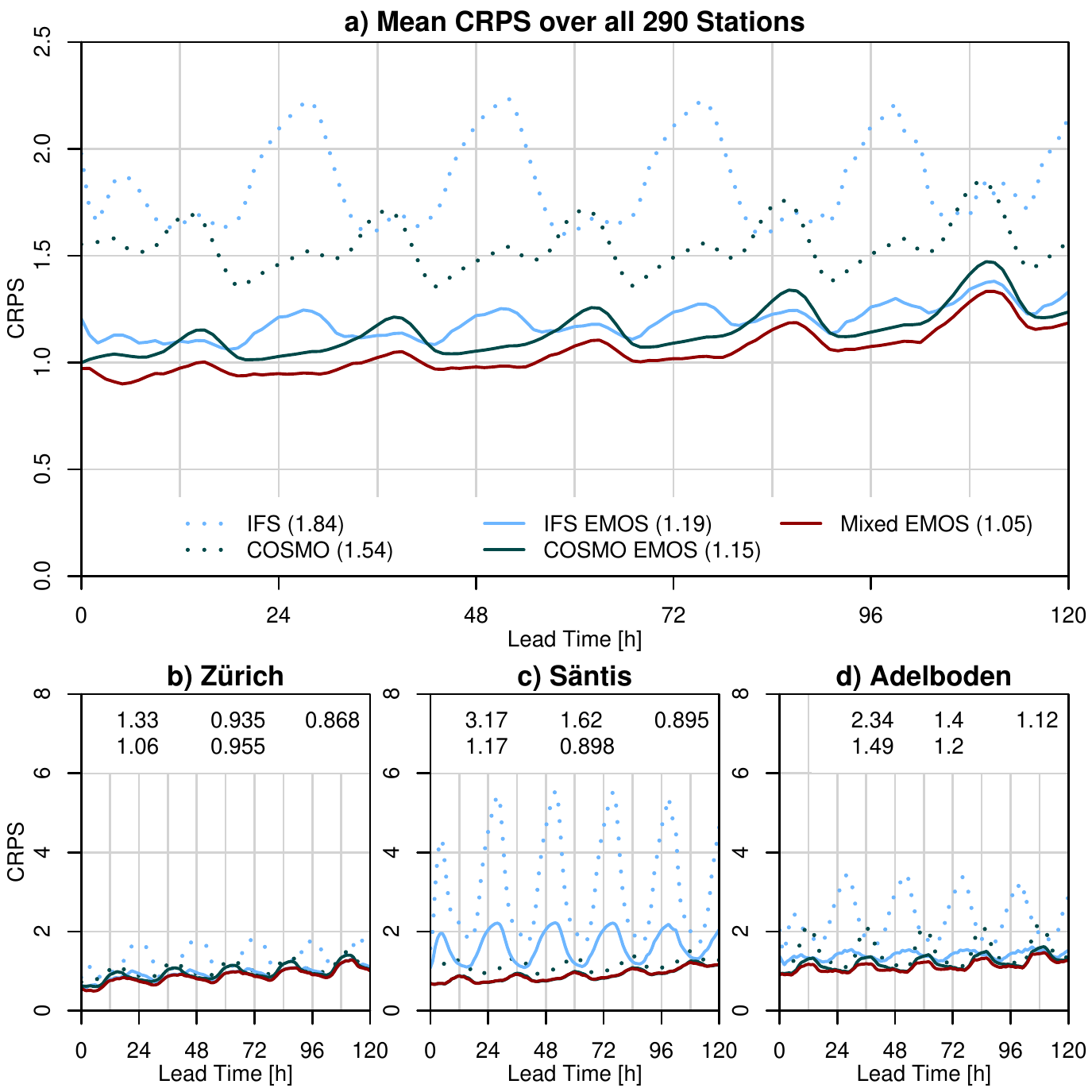}}
  \caption{Time series of the mean CRPS for the elevation-corrected direct model outputs (COSMO and IFS), the postprocessed forecasts (COSMO EMOS and IFS EMOS), and the Mixed EMOS. a) shows averages over all stations, and b) shows results for Z\"urich (lowland), c) for S\"antis (mountain top), and d) for Adelboden (valley floor). The values in brackets of a) are averages over the first 120 lead times; the values in b) - d) are in the same order.}
  \label{fig:CRPS}
\end{figure*}

Considering elevation-corrected model output, both models show a distinct but different diurnal cycle in forecast quality as measured by the CRPS (Fig. \ref{fig:CRPS}a). The discrepancy stems from a difference in the models' bias of the 2-m air temperature. IFS forecasts too low minimum temperatures and (too) high maxima, while COSMO has the opposite problem, hence its diurnal cycle of temperature is too weak. EMOS removes most of the bias, which is part of the reason why the CRPS is lower and the diurnal cycle of the score weaker. The diurnal cycle of the EMOS forecasts matches the observations much closer but both EMOS forecasts have a tendency for too high maximum temperatures (especially in summer) and minimum temperatures (particularly in winter). The other influence on the CRPS is the spread of the forecast, which also displays a diurnal cycle (not shown). The standard deviation of both models is too small, particularly for the first few days and for COSMO, therefore EMOS increases the standard deviation. The IFS EMOS has a larger spread in the night than COSMO while COSMO has a slightly larger standard deviation in the early afternoon, which could explain part of the remaining diurnal cycle in CRPS.
\\

\begin{figure*}[t]
  \centerline{\noindent\includegraphics[width=33pc,angle=0]{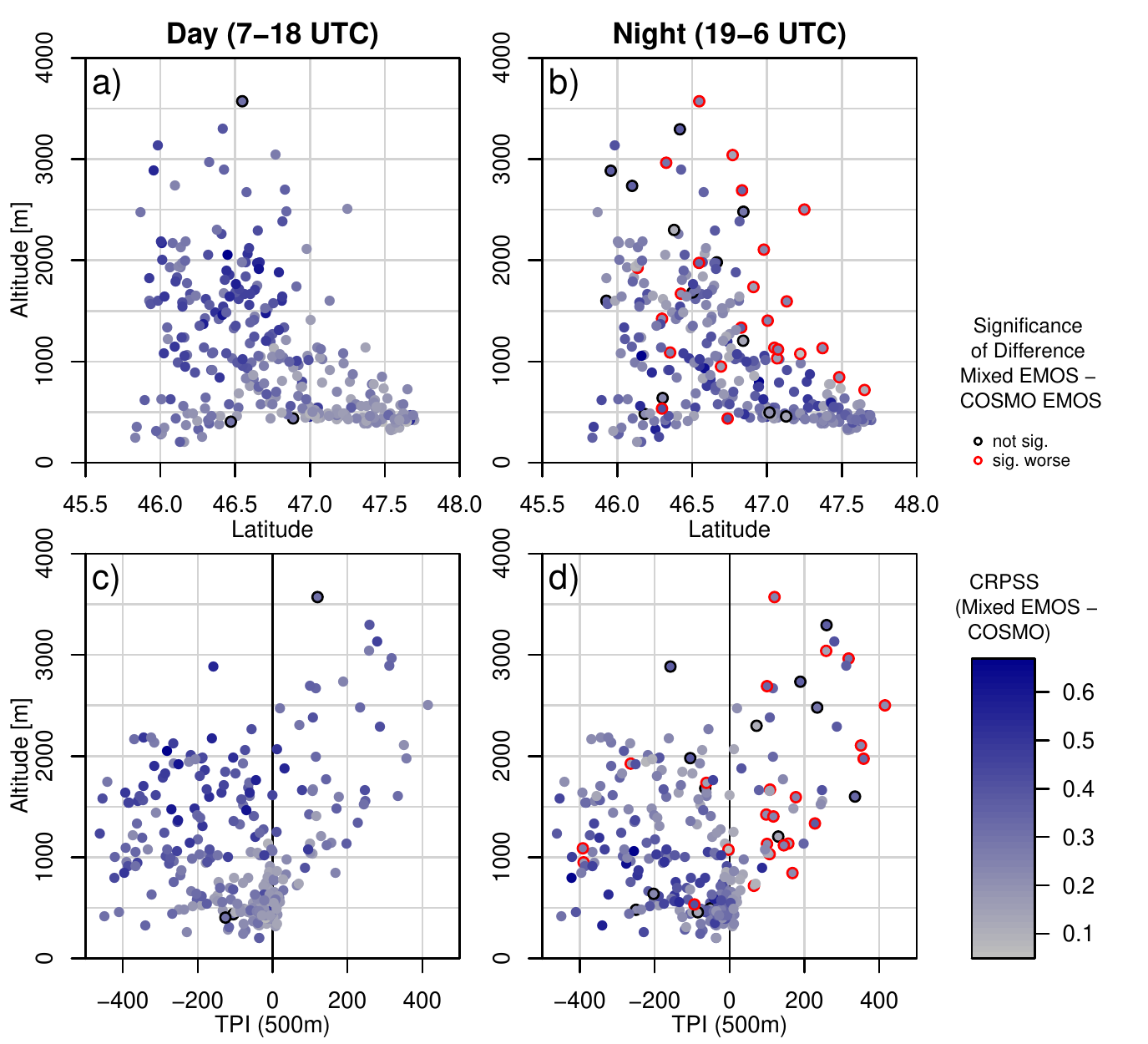}}
  \caption{Analysis of how the CRPSS of the Mixed EMOS depends on (a,b) latitude vs. elevation of the station, and (c,d) TPI vs. elevation. Blue-grey shading indicates the difference of the CRPSS of the Mixed EMOS compared to elevation-corrected COSMO, averaged over lead times 0-120 h, for day and night. Outer circles denote significance of the difference in the CRPS between Mixed EMOS and COSMO EMOS (significant level 0.05; red: Mixed EMOS is significantly worse, black: difference is not significant, no circle: Mixed EMOS is significantly better). All data is separated into day (a, c) and night (b, d).}
  \label{fig:day-night}
\end{figure*}

For the example of COSMO and IFS, it is shown that combining a higher- and coarser-resolution model enables a further improvement in forecast quality by 8-12\% (measured in terms of CRPSS) compared with the calibration of a single NWP. The average improvement amounts up to 30\% as compared to the high-resolution NWP model in operation (COSMO).
The benefit of the Mixed EMOS over IFS EMOS is especially large at high altitudes and peak locations, where EMOS alone is insufficient to correct the poor performance of IFS (see also Fig. \ref{fig:CRPS}c).
The improvement of Mixed EMOS over COSMO EMOS is generally smallest on mountain tops and at night (see Fig. \ref{fig:day-night}).
At night, peak locations along the northern slopes of the Alps have a significantly worse score for the Mixed EMOS forecast compared to COSMO EMOS. These exposed locations reach into the free atmosphere that is decoupled from the boundary layer at night, hence experiencing a smaller diurnal cycle in temperature.
In IFS, these places are located at too low altitude and the model cools too much at night. COSMO is better able to distinguish ridges and peaks and can better represent local-scale phenomena. Hence, the addition of IFS in exposed locations can be disadvantageous.
The largest improvement of Mixed EMOS over COSMO EMOS can be seen in the valleys, where even the high-resolution COSMO is insufficient to capture all relevant local-scale processes. The strong cooling of IFS at night tends to be more realistic in clear nights and therefore improves Mixed EMOS forecasts. 
In short, Mixed EMOS has a smaller bias and a smaller spread than either single-model EMOS in all seasons, leading to an improved score.
\\

The multi-model combination with Mixed EMOS allows to combine forecasts with a different forecast time horizon. In order to provide a seamless prediction across the transition from tow to one available ensemble, we proposed two simple approaches: One approach decreases the weight of the shorter NWP forecast in its last few timesteps, while the other approach is based on an extrapolation of the shorter NWP forecasts beyond its time horizon. While both approaches are fairly similar in terms of smoothness, the second approach is favorable as its overall skill is higher. Both approaches are motivated by the fact that we try to provide a smooth forecast without breaks or inconsistencies. The two approaches applied are straightforward and illustrative, and more sophisticated approaches can be developed in in future studies. 
There is also the potential to smooth additional seams in the proposed framework. Most notably, one with respect to observations potentially leading to improvements in forecast skill in the nowcasting time horizon (not shown), as well as with respect to monthly, seasonal and decadal forecasts and climate projections.
\\

\begin{figure}[t]
  \noindent\includegraphics[width=19pc,angle=0]{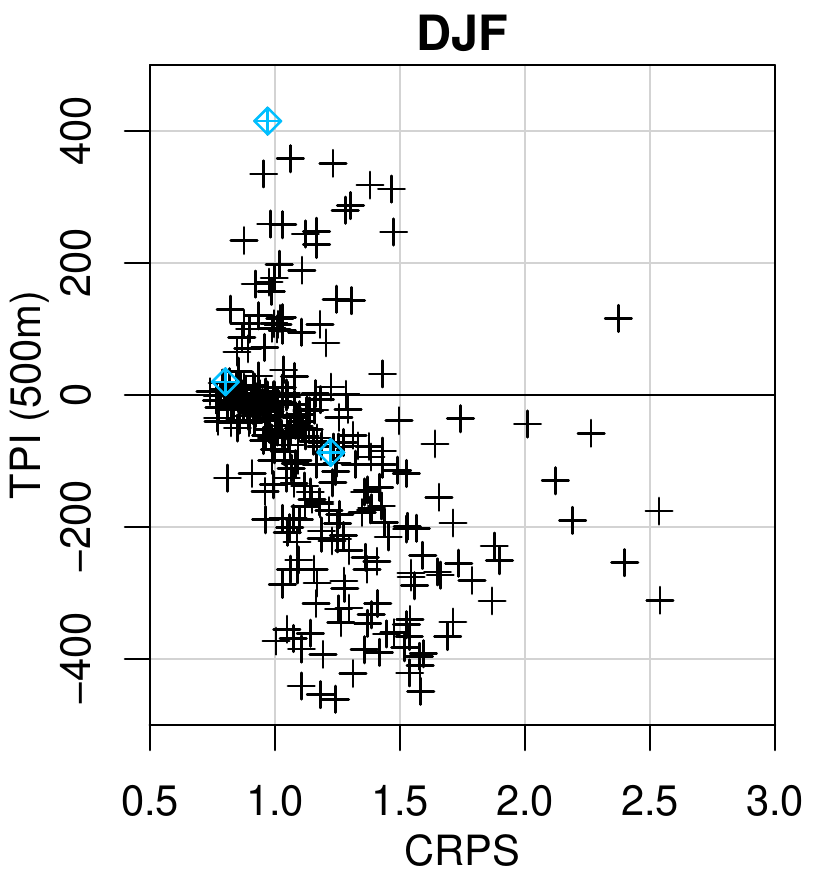}\\
  \caption{CRPS, averaged over lead times 0-120 h of the Mixed EMOS in winter (DJF) against TPI at 500 m resolution. The blue marked stations are (from top to bottom) S\"antis (mountain top), Z\"urich (lowland), and Adelboden (valley floor).}
  \label{fig:DJF}
\end{figure}

The remaining error of the Mixed EMOS depends on the season (Table \ref{tab:scores}). Winter (DJF) has the highest CRPS despite the second largest improvement in skill over COSMO. The pattern of the CRPS is related to TPI in winter (and fall) as shown in Fig. \ref{fig:DJF}. Locations with flat surroundings have the best score while stations within topography have worse CRPS values. In the lowlands, where the forecast is already good to begin with, postprocessing and model combination improves it even more. In the mountainous areas, the benefit of EMOS is largest but the error remains higher than in lowland areas. In a few cases, the CRPS value is still larger than 2. These stations are located in mid-altitude valleys (approx. 1000-2000 m), which are prone to the formation of cold air pools. (The one location in Fig. \ref{fig:DJF} with a positive TPI and high CRPS is situated in a small basin above a large valley, making the TPI at 500 m unsuitable to describe this location.) In clear winter nights, temperature at these locations can drop dramatically, which is hard to predict by NWP and only partly corrected by EMOS.
\\

The model combination in this study had ideal conditions that both models were available from the same initialization, i.e., the 00 UTC run from COSMO could be combined with the 00 UTC run from IFS. In operation, this is not always the case. Because IFS forecasts are available at a later time than forecasts form COSMO with the same initialization time, the newest available IFS is most often six hours older than COSMO, sometimes also 12 hours. We tested our Mixed EMOS with 12-hours older IFS, which resulted in only a minimal decrease in skill. Hence, Mixed EMOS will be a useful approach in operational use.
\\

Post-processing with EMOS is an appealing tool as it is able to substantially improve forecast accuracy and resolution, especially across complex topography. The framework offers flexibility to include additional predictors that may help further improving forecasts. These additional predictors can be static or model variables. Initial tests with the inclusion of TPI (a static predictor), or the predicted boundary layer height both showed promising results.
\\

The proposed multi-model combination presented in this study is attractive for several reasons. It improves forecast skill compared to single-model post-processing, particularly in valleys. It also allows to combine high-resolution NWP products, which usually exhibit frequent update cycles, with coarser-resolution NWP products, which are updated less frequently but provide predictions for longer lead times into a seamless forecast. A multi-model combined prediction thus profits from frequent updates of new forecast information that is of relevance in situations with low predictability (e.g., frontal passages and summertime convection), as well as from covering medium-range lead times (i.e., in the present example 15 days). The Mixed EMOS thereby generates medium-range forecasts in a seamless manner with improved skill during the period of overlap (i.e., up to five days in the present case). Additionally, the proposed multi-model combination offers more reliable operations as the risk of outage in operation is spread across multiple model sources. Finally, the Mixed EMOS approach is flexible for extensions to even more than two models or additional physical predictors. 

%
\datastatement
The observation data used in this work are available after registration on the data portal of MeteoSwiss (https://gate.meteoswiss.ch/idaweb). Model data are available from the authors upon request.

%
\acknowledgments
We acknowledge the following institutions for providing observations from their monitoring networks: Federal Office for the Environment (FOEN), numerous Cantons of Switzerland, MeteoGroup Schweiz AG, and the Swiss Federal Institute for Forest, Snow and Landscape Research (WSL). We thank Pirmin Kaufmann and Lionel Moret of MeteoSwiss for the discussion on the scores of different models and their comments on the manuscript.

%






%
%
%
\bibliographystyle{ametsoc2014}
\bibliography{references.bib}

%

\begin{table*}[t]
    \caption{Overview of the model data.}
    \label{tab:models}
	\centering
	\begin{tabular}{ l | c | c }
		\hline
		\bfseries{Model} & \bfseries{COSMO-E} & \bfseries{IFS-ENS} \\ \hline \hline
		\bfseries{Name} & COnsortium for Small-scale MOdeling - Ensemble & Integrated Forecasting System - ENSemble \\ \hline
		\bfseries{Operated by} & MeteoSwiss & ECMWF \\ \hline
		\bfseries{Ensemble Members} & 21 & 51 \\ \hline
		\bfseries{Domain} & regional & global \\ \hline 
		\bfseries{Horizontal Resolution} & 2.2 km & 18 km \\ \hline
		\bfseries{Vertical Levels} & 60 & 91 \\ \hline
		\bfseries{Lead Time} & 120 h (5 days) & 360 h (15 days) \\ \hline
		\bfseries{Temporal Resolution} & 1 h & 1/3/6 h \\ \hline
		\bfseries{Initialization} & 2 per day & 2 per day\\ \hline
	\end{tabular}
\end{table*}

\begin{table*}[t]
    \caption{(Skill) scores averaged over lead times 0-120 h and over spring (MAM), summer (JJA), fall (SON), winter (DJF) and all year (ANN). Top: The CRPS of the elevation-corrected direct model outputs (COSMO and IFS), and postprocessed forecasts (COSMO EMOS and IFS EMOS), and the Mixed EMOS. Middle: The CRPSS of the same forecasts relative to the elevation-corrected COSMO. Bottom: The percentage of stations which have a positive CRPSS.}
    \label{tab:scores}
    \centering
    \begin{tabular}{c  c | c | c | c | c | c}
        \hline
         & & \bfseries{COSMO} & \bfseries{IFS} & \bfseries{COSMO EMOS} &    \bfseries{IFS EMOS} & \bfseries{Mixed EMOS} \\ \hline \hline
        \multirow{5}{*}{\bfseries{CRPS}} & MAM & 1.64 & 1.90	& 1.20 & 1.19 & 1.06 \\ \cline{2-7}
         & JJA & 1.40 & 1.48 & 1.05 & 1.05 & 0.971 \\ \cline{2-7}
         & SON & 1.41 & 1.69 & 1.10 & 1.18 & 1.01 \\ \cline{2-7}
         & DJF & 1.73 & 2.34 & 1.26	& 1.40 & 1.17 \\ \cline{2-7}
         & ANN & 1.54 & 1.84 & 1.15	& 1.19 & 1.05 \\ \hline \hline
        
        \multirow{5}{*}{\bfseries{CRPSS}} & MAM & -- & -0.210 & 0.234 & 0.227 & 0.320 \\ \cline{2-7}
         & JJA & -- & -0.113 & 0.220 & 0.215 & 0.279 \\ \cline{2-7}
         & SON & -- & -0.243 & 0.201 & 0.133 & 0.259 \\ \cline{2-7}
         & DJF & -- & -0.395 & 0.252 & 0.162 & 0.307 \\ \cline{2-7}
         & ANN & -- & -0.231 & 0.241 & 0.204 & 0.306 \\ \hline \hline
         
        \multirow{5}{*}{\parbox{1.5cm}{\bfseries{Stations CRPSS > 0 [\%]}}}& MAM & -- & 34.5 & 100 & 88.6 & 100 \\ \cline{2-7}
         & JJA & -- & 50.3 & 98.6 & 85.9 & 99.7 \\ \cline{2-7}
         & SON & -- & 37.9 & 99.0 & 73.4 & 100 \\ \cline{2-7}
         & DJF & -- & 23.4 & 100 & 82.4 & 100 \\ \cline{2-7}
         & ANN & -- & 33.1 & 100 & 90.3 & 100 \\ \hline
    \end{tabular}
\end{table*}

%

\end{document}